\documentclass[conference]{IEEEtran}
\IEEEoverridecommandlockouts
\usepackage{cite}
\usepackage{amsmath,amssymb,amsfonts}
\usepackage{algorithmic}
\usepackage{graphicx}
\usepackage{textcomp}
\usepackage{xcolor}
\usepackage{array}
\usepackage{cite}
\pdfoutput=1

\def\BibTeX{{\rm B\kern-.05em{\sc i\kern-.025em b}\kern-.08em
    T\kern-.1667em\lower.7ex\hbox{E}\kern-.125emX}}
\begin{document}

\title{Fundamental Electron and Potential Relations in Dilute Plasma Flows}

\author{\IEEEauthorblockN{Shiying Cai, Chunpei Cai, Xin He}
\IEEEauthorblockA{\textit{Dept. of Mechanical Engineering-Engineering Mechanics} \\
\textit{Michigan Technological University}\\
Houghton, MI 49931, USA \\
David L. Cooke \\
Space Vehicle Directorate, AFRL, Kirtland Air Force Base \\
2000 Wyoming Blvd SE, Albuquerque, NM 87123 \\
Email:ccai@mtu.edu}
}

\maketitle
\thispagestyle{plain}
\pagestyle{plain}

\begin{abstract}
In this short note, we present some work on investigating electron temperatures and potentials in steady or unsteady dilute plasma flows.  The analysis is based on the detailed fluid model for electrons. Ionization, normalized electron number density gradients, and magnetic fields are neglected. The transport properties are assumed as local constants. With these treatments, the partial differential equation for electron temperature degenerates as an ordinary differential equation.  Along an electron streamline, fundamental formulas for electron temperature and plasma potential are obtained. These formulas offer significant insights, 1). for steady flow, the electron temperature and plasma potential distributions along an electron streamline include two exponential functions, and the one for plasma potential includes an extra linear distribution function; 2). for unsteady flows, both the temporal and spatical parts include potential functions. 
\end{abstract}


\section{Introduction}
In dilute plasma flows, the continuum flow assumption breaks down, and there are multiple species, multiple time and space scales. For light electrons, the oscillation frequency and speeds are quite high, while heavy ions and neutrals have slow motions. For many situations, multiple physics must be considered, e.g., electric and magnetic fields, rarefication, and collisions.  Investigations on plasma flows are usually quite challenging, and many assumptions may be needed in order to perform theoretical, numerical and experimental studies.

Numerical simulations of dilute plasma are quite helpful to understand the related physics, and there are several simulation approaches. The first approach uses full scale Particle-In-Cell (PIC) \cite{PIC} simulations, where electrons, ions, and neutrals are treated as different particles. However, these full scale PIC simulations track each movement of each particle, with too many unnecessary details because for many situations, only the collective behaviors of different species are of special concerns. The simulation time steps in PIC simulations are limited by the periods of gyro-motion of electrons, ions, the mean collision time, and the plasma frequency. Due to these reasons, full-scale PIC simulations are quite expensive and not suitable to simulate many practical engineering problems. The second approach solves the magneto-hydro-dynamic equations\cite{chen}, where ions and electrons are treated as a mixture. This method solves for the macroscopic properties, and it cannot offer separated information for electrons and ions. Another practical approach for plasma simulation is a hybrid method \cite{yim} which simulates heavy ions and neutrals with the direct simulation Monte Carlo (DSMC) \cite{DSMC} and the PIC methods; electrons are treated as fluids due to their higher mobility. Compared with methods of solving ions and electron as  two fluids\cite{two}, this hybrid treatment provides detailed collective information for electrons and a certain degree of accuracy are maintained, while the simulation speed is much faster than full scale PIC simulations. In many applications, the hybrid method is effective and sufficiently accurate. Other simulation approaches include the directly solving \cite{aristov} the Vlasov equation. However, numerical simulation results are usually case by case.

Experimental studies on plasma flows are quite important and sometimes un-replaceable because they can validate the theoretical and simulation results. However, related experiment costs can be high. By comparison, theoretical investigations and modelling can be quite helpful and can offer some deep insights.

For plasma flows, the electron temperature is one important property. Other than the electron number density and bulk velocity components, electron temperature describes the thermal motion behaviors and can reveal much information for plasma flows. For example, dilute plasma flows from a Hall effect electric propulsion (EP) device can be used for thruster performance evaluations and contamination estimations, and electron temperature can be measured conveniently. The electron thermal temperature and potential increases  across the pre-sheath in front of a Tethered- Satellite-System can result in significant current enhancement to the satellite\cite{Cooke}. It is also reported that new differential equations can be developed to describe the heat fluxes into the wall with two-temperature modelling of the thermal plasmas. Due to the significant temperature differences between the electrons and ions, usually electrons deposit significant energy into plasma (e.g. \cite{pekker,weifz}).

In this paper, we present some recent modelling work on analyzing electron temperatures and plasma potentials in both steady and unsteady dilute and cold pasma flows. The approach is based on the fluid model for electrons, and the results are semi-analytical. The number density gradients in the dilute plasma flows are assumed to be small, the magnetic field effects are also assumed negligible.

The remaining sections are organized as follows. Section \ref{sec:detailed} explains the detailed electron model, section \ref{sec:steady} reports formula derivations the relations between electron temperatures and plasma potentials, and section \ref{sec:unsteady} presents the corresponding results for the situation of  unsteady flow situations, and in the end, section \ref{sec:conclusions} summarizes and concludes this paper.

\section{Detailed Fluid Model For Electrons}
\label{sec:detailed}
The continuity equation for electrons is:
\begin{equation}
  \frac{\partial n_e}{\partial t}  + \nabla \cdot  (n_e \overrightarrow{v_e})   = \dot{S_e},
\label{eqn:continuity}
\end{equation}
where $n_e$ and $\overrightarrow{v_e}$ is the electron number density and electron velocity vector, $\dot{S_e}$ is the local electron generate rate.  If the magnetic field is neglected, the generalized Ohm's law states:
\begin{equation}
   \overrightarrow{j}  = \sigma \left(  - \nabla \phi + \frac{1}{en_e} \nabla (n_e k T_e) \right),
\label{eqn:j}
\end{equation}
where $\overrightarrow{j}$ is electricity current, $\phi$ is the plasma potential, $e$ is the unite charge for electrons, $\sigma$ is electricity conductivity, $T_e$ is the electron temperature, and $k$ is the Boltzmann constant.

The charge continuity condition is:
\begin{equation}
    \nabla \cdot \overrightarrow{j}  =0.
\label{eqn:con_den}
\end{equation}
The electron  temperature equation can date back to the 70s'\cite{Mitcher}:
\begin{equation}
\begin{array}{rll}
&\frac{\partial}{\partial t} \big(\frac{3}{2} k T_e n_e  + \frac{1}{2} m_e v_e^2  n_e\big)  + \nabla \cdot \big(   \frac{3}{2} k T_e n_e  \overrightarrow{v_e}  + 
  & \\ 
  &\frac{1}{2} m_e v_e^2 n_e \overrightarrow{v_e}\big)  + \nabla \cdot \overrightarrow{q_e} - \overrightarrow{j} \cdot \overrightarrow{E} + \dot{E_e}  + \frac{3}{2} k n_e   &\\
&\big( \frac{2m_e}{m_e+m_h} \nu_e (T_e -T_h) \big)  + p_e (\nabla \cdot \overrightarrow{v_e})=0 &
\end{array}
\label{eqn:orginalTe}
\end{equation}
where $m_e$ and $m_h$ are the mass for electron and heavy particles, $\overrightarrow{q}_e$ is the electron heat flux,  $\overrightarrow{E}$ is the electricity field vector,  $\nu_e$ is the electron collision rate, $T_h$ is heavy particle temperature, which is assumed as a constant, and $p_e$ is the electron pressure.  The above relation can be simply understood from the first law: time change rate of the electron total energy (the 1st term at the left hand side) + convection of energy (the 2nd term)  + energy conduction via the Fourier's law (the 3rd term) + Joule heating (the 4th term) + electron energy loss due to ionization (the 5th term, $\dot{E_e}$)+ allocations of translational energy between heavy particles (neutrals or ions) and electrons (the 6th term). In general, the electron kinetic energy $\frac{1}{2} m_e v_e^2 n_e$ can be assumed negligible than the thermal energy $\frac{3}{2} k T_e n_e$ and removed from the  energy relation.  The Fourier's heat transfer law  states $\overrightarrow{q_e}= - \kappa_e(T_e) \nabla T_e$, where $\kappa_e(T_e)$ is the electron conduction coefficient. Here the electron conductivity is assumed to be a function of electron temperature. Then, $ \nabla \cdot \overrightarrow{q_e}  = -\nabla \cdot (\kappa_e(T_e) \nabla T_e) = -\kappa_e (T_e) \nabla^2 T_e - \nabla \kappa_e(T_e) \cdot \nabla T_e$.

\section {Steady state relations}
\label{sec:steady}
In this section, the plasma flow is assumed to be steady and the time change rate of the electron total energy is neglected.  After simple derivations, the following partial differential equation is obtained:
\begin{equation}
\begin{array}{rll}
  \nabla^2 T_e  = & - \nabla \mbox{ln} (\kappa_e) \cdot \nabla T_e  + \frac{1}{\kappa_e} \bigg(  - \overrightarrow{j} \cdot \overrightarrow{E}  +  & \\
  &\frac{3}{2} n_e \overrightarrow{v_e} \cdot  \nabla( kT_e)  + \frac{3}{2} kT_e \nabla \cdot (n_e \overrightarrow{ v_e})   &  \\
    &+ 3 \frac{m_e}{m_h} \nu_e n_e k (T_e -T_h) +  \dot{E_e}  +p_e (\nabla \cdot \overrightarrow{v_e}) \bigg) &
\end{array}
\label{eqn:Te}
\end{equation}

To further simplify the above relations for the electron temperature distribution,  the following assumptions are adopted:

{\bf Assumption 1)}. The plasma flow is steady;

{\bf Assumption 2)}. The transport coefficients and electric conductivity $\sigma$, are constant through the whole flowfield as a first order approximation;

{\bf Assumption 3)}. The normalized electron number density gradient, $\frac{1}{n_e} \nabla n_e$ is small; hence, Eqns. \ref{eqn:j} and \ref{eqn:con_den} degenerate to:
\begin{equation}
         \nabla^2 \phi =\frac{k}{e} \nabla^2 T_e,
\end{equation}
A simple integration leads to the following relation:
\begin{equation}
   \nabla \phi  =  \frac{k}{e} \nabla T_e  + \overrightarrow{a};
\label{eqn:phiT}
\end{equation}
where $\overrightarrow{a}$ is a constant vector for the whole flowfield, and it is to be determined later.

{\bf Assumption 4)}. Ionization effects can be neglected, i.e., the last term in Eqn.\ref{eqn:Te} is zero.  Also, Eqn.\ref{eqn:continuity} degenerates  as  $\nabla \cdot (n_e  \overrightarrow{v_e} ) = 0$  along electron streamlines, for the whole steady flowfield;  and

{\bf Assumption 5)}. Because the large difference between the temperature for electrons and the temperature for ions or neutrals, $T_h$ is considered as a constant in the term for translational energy re-allocation after collisions. Here the subscript $_h$ represents heavy particles including ions and neutrals. 

{\bf Assumption 6)}. The Joule heating term
\[
 \overrightarrow{j}\cdot \overrightarrow{E} = \sigma [-\nabla \phi + \frac{k}{e} \nabla T_e ] \cdot [-\nabla \phi] = \sigma a_s^2 + \frac{\sigma k}{e} a_s  \nabla_s T_e,
\] where subscript ``s'' represents a component along a streamline. 

Assumptions 3 and  4 lead to $\nabla \cdot (n_e \overrightarrow{v_e} ) = 0  =  n_e  \nabla \cdot  \overrightarrow{v_e}   + \overrightarrow{v_e} \cdot \nabla n_e$,  and  $\nabla \cdot  \overrightarrow{v_e} =0$.  Then the work done by expansion, $p_e (\nabla \cdot \overrightarrow{v}_e)=0$, can be neglected.

With the above assumptions, Eqn.5 degenerates as:
\begin{equation}
        \nabla^2 T_e  +  \overrightarrow{A_0} \cdot \nabla T_e  +  A_1 (T_e-T_h) +A_2 =0;
\label{eqn:TeOde}
\end{equation}
where
\begin{equation}
\begin{array}{rll}
    \overrightarrow{A_0} &=&  \frac{1}{\kappa_e} ( \frac{\sigma k}{e} \overrightarrow{a} - \frac{3}{2} n_e \overrightarrow{v_e} k + \nabla \kappa_e), \\
                   A_1   &=& -\frac{3}{\kappa_e} \frac{m_e}{m_h} \nu_e n_e k,  A_2   = \frac{\sigma a^2}{\kappa_e},
\end{array}
\end{equation}
Obviously, $A_1<0$ and $A_2>0$.

In general, Eqn.\ref{eqn:TeOde} is difficult to handle with multiple varying coefficients and multi-dimensions. As an initial effort, this work discusses the one-dimensional scenario by concentrating on properties along a streamline, then there is a general solution to this differential equation, coefficients $\overrightarrow{A}_0$, $A_1$ and $A_2$ are assumed as local constants:
\begin{equation}
\begin{array}{rll}
    T_e(s)  &=& T_h - \frac{A_2}{A_1} +C_1  e^{\lambda_1 s} + C_2  e^{\lambda_2 s}  \\
            &\equiv& C_0 +C_1  e^{\lambda_1 s} + C_2 e^{\lambda_2 s};   \\
            & & \lambda_{1,2} = -\frac{A_0}{2} \pm \frac{1}{2} \sqrt{A_0^2 - 4A_1}
\end{array}
\label{eqn:Tesoultion}
\end{equation}
Obviously, $\lambda_1$ is positive and $\lambda_2$ is negative; $C_0 = T_h- A_2/A_1$, $C_1$ and $C_2$ are two constants to be determined with proper boundary conditions. For certain, along the streamline, one exponential function decays and the other increases. Evidently, $\lambda_{1,2}$ values, and those coefficients $C_0$, $C_1$, and $C_2$ must adjust along the streamline, to maintain bounded electron temperature. 

Correspondingly,  the plasma potential along a streamline is:
\begin{equation}
    \phi(s) =  \frac{k}{e} ( C_1  e^{\lambda_1 s}  +  C_2 e^{\lambda_2 s} ) + a_ss +b
\label{Eqn:potential}
\end{equation}
where and $b$ is  a constant to be determined with boundary conditions.

It shall be mentioned that in the literature there are other models for the electron temperature (e.g.\cite{Aleksandrov}); however, Eqn.\ref{eqn:orginalTe} is the most convenient one for development due to the simple near linear format. This advantage  has not been fully explored. Some past work adopted Eqn.\ref{eqn:orginalTe}, but magnetic field (e.g. \cite{Kawamura}) is also involved and hence the expressions  are rather complex for further simplifications.

Numerical simulations of steady plasma flows from two Hall thrusters were performed \cite{Cai0,Cai1,Cai2} with the hybrid simulation method. For the first test case, along the thruster centerline, the near-field potentials and electron temperatures were close to the experimental measurements. The profiles for electron temperatures and plasma potentials at further downstream were constructed by using detailed information at one upstream point, and the values at another downstream point. The profiles develop as exponential functions and were close to simulation results. For the second test case, the electron temperature and plasma potential profiles along the centerline were approximated by using double-exponential functions. In general, the comparisons were satisfying, but more investigations are further needed in the future, for example, more specific solutions for plasma flows of different dimensions, and possible new simulations with different simulation methods.

\section{Unsteady state solutions}
\label{sec:unsteady}
In the above, two semi-analytical solutions for steady, dilute, cold plasma flows have been  obtained. One is for the plasma potential, and the other is for electron temperature. These two relations illustrate physical factors and their contribution to the steady electron flow properties. The relations hold along electron streamlines. 

If the steady flow and small density gradient assumptions in the above work are further relaxed,  it can be demonstrated that there still exists semi-analytical solutions for the plasma potential and electron temperature.

For unsteady flow, Eqn.\ref{eqn:orginalTe} becomes:
\begin{equation}
\begin{array}{rll}
& \frac{3}{2} k  n_e \frac{\partial T_e}{\partial t}  + \frac{3}{2}  k n_e  v_e \frac{\partial T_e}{\partial s}  -\kappa_e \frac{ \partial^2 T_e}{\partial s^2} - \frac{\partial \kappa_e}{\partial s} \frac{\partial T_e}{\partial s}  \\
&    - \overrightarrow{j} \cdot \overrightarrow{E}  + 3 k n_e  \frac{m_e}{m_h} \nu_e T_e +n_e kT_2 (\nabla \cdot \overrightarrow{v_e}) =0
 \end{array}
\label{eqn:newenergy}
\end{equation}

or, 
\begin{equation}
     D_0 \frac{\partial T_e}{\partial t}  + D_1 \frac{\partial T_e}{\partial s}  - \kappa_e \frac{ \partial^2 T_e}{\partial s^2} + D_3 (T_e  + D_4) =0
\label{eqn:master}
\end{equation}
where  $D_0 = \frac{3}{2} k  n_e$, $D_1 =D_0 v_{es} - \frac{\partial \kappa_e }{\partial s} - \frac{\sigma k}{e} a_s$,   $D_3 =3 kn_e \nu_e \frac{m_e}{m_h}$,  and  $D_4  =-( \sigma a_s^2 + T_h)/D_3$. The Joule heating expression in included. 

By introducing  a new variable $T(s,t) \equiv T_e +D_4  \equiv X(t)Y(s)$,  Eqn. \ref{eqn:master} transforms as:
\begin{equation}
  D_0 \frac{X'(t)}{X(t)} + D_1 \frac{Y'(s)}{Y(s)} -\kappa_e  \frac{Y''(s)}{Y(s)} + D_3 =0
\label{eqn:m1}
\end{equation}
With the method of variable separation, the above equation leads to the following solution:
\begin{equation}
\begin{array}{rll}
   X(t) &=& B_1 e^{ D_6 t}, Y(s) = B_2 e^{r_1 s} + B_3e^{r_2 s}, \\
   r_{1,2} &=& \left(  D_1  \pm \sqrt{ D_1 ^2   + 4 \kappa_e D_5} \right)/(2\kappa_e)
\end{array}
\end{equation}
where  $D_5$, $B_1$, $B_2$, $B_3$ and $D_6 = (D_5-D_3)/D_0$  may be determined later  with boundary  and initial conditions. The final expression for $T_e$ is:
\begin{equation}
  T_e(s,t) = B_1 e^{D_6t} \left[ B_2 e^{r_1s}+B_3 e^{r_2s} \right]  - D_4 
\label{eqn:unsteady_Tesoultion}
\end{equation}
As shown, there may be oscillating wave behaviors in this distribution because  $r_1, r_2$ and $D_6$ can have imaginary components.  Eqns. \ref{eqn:unsteady_Tesoultion} and \ref{eqn:Tesoultion} are completely compatible, for steady flows, the time term in Eqn.\ref{eqn:unsteady_Tesoultion} disappears, and the equation automatically degenerates to Eqn. \ref{eqn:Tesoultion}.

The Maxwell's equations \cite{chen} are adopted in this investigation:
\begin{equation}
  \frac{ \partial \overrightarrow{B} }{\partial t}  = - \nabla \times \overrightarrow{E};
           \nabla \times \overrightarrow{B}  =    \mu_0  \left(  \overrightarrow{j}   +  \epsilon_0    \frac{ \partial \overrightarrow{E} }{\partial t}   \right)
\label{eqn:max1}
\end{equation}
The magnetic field factor can be eliminated by combining the above two relations. The following equation is obtained:
\begin{equation}
   - \nabla  \times  (\nabla  \times  \overrightarrow{E})  = \mu_0   ( \frac{\partial{\overrightarrow j}}{ \partial t}  + \epsilon_0   \frac{ \partial^2 \overrightarrow E}{ \partial t^2 } )
\label{eqn:gov1}
\end{equation}
In the above equation, the left hand side term is zero due to the fact that $\overrightarrow{E} =-\nabla \phi$. The generalized Ohm's law, or Eqn.\ref{eqn:j} transforms into  one partial differential equation:
\begin{equation}
  -  \frac{ \partial^2 \phi  }{  \partial t \partial s}  + \frac{\partial}{\partial t} \left(\frac{1}{en_e} \nabla_s (n_e kT_e) \right)    - \frac{ \epsilon_0 }{\sigma}   \frac{\partial^3 \phi }{\partial t^2 \partial s}   =0
\label{eqn:cur_1}
\end{equation}
The second term in the above equation becomes:
\begin{equation}
\frac{1}{en_e}  \nabla_s  (n_e kT_e)  =  \frac{kT_e}{e}   \nabla_s ( ln n_e )  + \frac{k}{e} \nabla_s T_e
\label{eqn:f1}
\end{equation}
Because the first term is highly nonlinear, relation $n_e \approx n_{ref}$ is adopted and this term is neglected.  Then, Eqn. \ref{eqn:cur_1} becomes:
\begin{equation}
  -\frac{ \partial^2 \phi }{ \partial t \partial s}   + \frac{k}{e}  \frac{\partial^2  T_e}{\partial t \partial s}  -  \frac{\epsilon_0}{\sigma}   \frac{\partial^3 \phi}{\partial s \partial t^2} =0
\end{equation}

After integrating twice over $s$ and $t$, the following expression is obtained:
\begin{equation}
 -\frac{\epsilon_0}{\sigma}  \frac{\partial \phi}{ \partial t}  + \frac{k}{e} X(t)Y(s) - \phi  +Ps +Qt +L  =0
\end{equation}
where  $P$ and $Q$ must be zero to maintain finite values for the above equation, at farfield and as time with large $s$ and $t$ increases to infinity.  $L$ can be combined into $\phi$, as $p(s,t) = \phi-L$, which can  be defined as two parts $p(s,t) = a(t)b(s)$. Then, a new relation is created:
\begin{equation}
  \frac{\epsilon_0}{\sigma} \frac{ a'(t)}{ a(t)}  +1  -\frac{k}{e}  \frac{ X(t) }{a(t)}     \frac{Y(s) }{b(s)} =0.
\end{equation}
As the other terms are functions of $t$ only,  $Y(s)/b(s)$ must be a constant which is defined as $K$, with a solution $b(s) =Y(s)/K$.  The above equation transforms as:
\begin{equation}
    \frac{\epsilon_0}{\sigma} a'(t) +a(t) - \frac{kB_1K}{e} e^{D_6 t} =0
\end{equation}
Then, the solution for $a(t)$ is:
\begin{equation}
    a(t) =B_3 e^{-\sigma t/\epsilon_0}  +B_4 e^{D_6t}, \hspace{0.3cm}  B_4 =\frac{kB_1K}{e} \frac{\sigma }{ \epsilon_0 D_6 +\sigma }.
\end{equation}
where $B_3$ is a constant. 

Finally,  the solution for the potential is:
\begin{equation}
\begin{array}{rl}
&\phi(s,t)= L + \left( B_3 e^{-\sigma t/\epsilon_0}  + B_4  e^{D_6 t} \right) \frac{Y(s)}{K} =L+ \\
& \frac{1}{K} \left( B_3 e^{-\sigma t/\epsilon_0}  + B_4  e^{ D_6 t} \right) (B_2e^{r_1 s} + B_3e^{r_2 s})
\end{array}
\label{eqn:unstady_pot}
\end{equation}
where  $K$, $B_2$, $B_3$, $B_4$ and $L$  are constants to be determined. 

Equation \ref{eqn:unstady_pot}  indicates the potential develop with oscillations in time and locations along the electron streamline.  Because the term $\sigma/\epsilon >0$, the first term in the first bracket  decays with time. However, the other coefficients in this expression can include imaginary parts, the plasma potential can grow, decay, or remain stable for both time and streamline, and the oscillations with sine or cosine profiles are highly possible for time and space. -As well known, oscillations are fundamental features in plasma flows. For this situation, the Joule heating term can include oscillations in space and time as well, and the amplitude can vary, its effect is not limited to dissipation.

\section{Conclusions}
\label{sec:conclusions}
A study on electron temperatures and plasma potentials in steady or unsteady cold and dilute plasma flows is performed. Started from the detailed fluid model for electrons and with several assumptions, compact exact formulas for electron temperatures and plasma potentials are obtained. The derivation is along the electron streamline. 

For the steady flow situation, all formulas include two exponential functions, and the formula for plasma potentials includes an extra linear distribution term. An assumption of local constant coefficients is adopted which is generally more reasonable at locations without large gradients, such as the farfield plume flows from a Hall effect thruster. The curves shall be very smooth.

For the unsteady flow situation, the situations are complex. There are still two components in the solutions to the electron temperature and plasma potential. One part is for the  time factor and the other for the streamline.  However, both parts may involve sine or  cosine functions, or exponential functions and wave patterns. The trends for both parts can also can be steady, decaying or amplifying.   

Numerical simulations are  need to be performed in the next step, several preliminary simulation results are already performed.\cite{Cai1, Cai2, unsteady} However, due to the many constants in those formulas, it is not possible to give very accurate agreement, instead, the simulation can only demonstrate related patterns.  


\section*{Acknowledgement}
This work was supported by AFRL with contract No. FA9550-15-F-0001.

\vspace{12pt}

\end{document}